\documentclass[a4paper,prl,twocolumn,floatfix,footinbib]{revtex4}

\usepackage{amsmath,amssymb}
\usepackage{graphicx}

\begin{document}

\title{Enhanced tunneling across nanometer-scale metal-semiconductor
interfaces}

\author{G.D.J. Smit}
\author{S. Rogge}
\author{T.M. Klapwijk}
\affiliation{Department of Applied Physics and DIMES, Delft
University of Technology, Lorentzweg 1, 2628 CJ Delft, The
Netherlands}

\date{\today}

\begin{abstract}
We have measured electrical transport across epitaxial,
nanometer-sized metal-semiconductor interfaces by contacting
CoSi$_2$-islands grown on Si(111) with the tip of a scanning
tunneling microscope. The conductance per unit area was found to
increase with decreasing diode area. Indeed, the zero-bias
conductance was found to be $\sim 10^4$ times larger than expected
from downscaling a conventional diode. These observations are
explained by a model, which predicts a narrower barrier for small
diodes and therefore a greatly increased contribution of tunneling
to the electrical transport.
\end{abstract}

\maketitle

Electrical transport through metal-semiconductor interfaces has
received tremendous interest in the past decades, both
experimentally and theoretically. Nevertheless, an important
shortcoming of existing models is the restriction to infinitely
extending interfaces, so that all parameters vary only in the
direction perpendicular to the surface. When the interface-size
enters the nanoscale regime, many of these models cease to apply.
Only a few experiments addressing this topic have been reported.
In none of them epitaxial interfaces were used. Scanning tunneling
spectroscopy (STS) of metallic clusters on a semiconductor surface
has been used to study small metal-semiconductor contacts
\cite{carroll97}. In addition, experiments have been carried out
in which the tip of a scanning tunneling microscope (STM) was used
to contact a semiconductor surface \cite{avouris93,hasunuma98} or
a metallic cluster on a semiconductor surface \cite{hasegawa99} to
form a small Schottky contact. Various deviations from the
large-diode models were revealed, e.g.~enhanced conductance, which
was interpreted as a lower effective barrier. Besides the work
that addresses a single small diode directly, measurements have
been carried out on many small diodes in parallel
\cite{doniach87,yang97}.

In this paper, we present measurements of electrical transport
through an epitaxial, nanometer sized metal-semiconductor
interface. We argue that the observations can be explained by a
simple model for the Schottky barrier thickness in
metal-semiconductor interfaces smaller than the free-carrier
screening length (Debye length, $L_{\mathrm{D}}$). Our model
predicts an interface-size dependent barrier thickness, leading to
greatly enhanced tunneling in small Schottky diodes. The
CoSi$_2$/Si(111)-interface used in our experiments is among the
few metal-semiconductor interfaces of which reliable Schottky
barrier height (SBH) values exist, mainly because it can be grown
as a virtually perfect, abrupt, epitaxial interface
\cite{tung92b}. The SBH in this system is 0.67~eV (for n-type Si)
and has been measured with various techniques
\cite{tung92b,kaiser91,kaenel00}. It is therefore a nearly ideal
system to study electrical properties of metal-semiconductor
interfaces and has been intensely used for that purpose.

Both in our model and in the analysis of the measurements, the SBH
will be considered as a given quantity, because of the
well-determined character of the CoSi$_2$-Si(111) interface. We do
not expect that ultra-small-size effects as reported in
Ref.~\onlinecite{landman00} play a role here, due to the
semi-infinite extension of the semiconductor in our experiment.
The focus will be on the size and shape of the space charge region
in the semiconductor. The resulting band bending gives in general
rise to a potential barrier in the semiconductor, the shape of
which is very important as it determines the conductance due to
the various transport mechanisms ({\frenchspacing e.g. thermionic}
emission, tunneling) across the interface.

In a large diode (an infinite metal-semiconductor interface) the
extent of the space charge layer in the semiconductor that
compensates the surface charge in the metal is set by the
concentration of free carriers and doping atoms. The width of this
space charge layer (and thus the thickness of the barrier) is
therefore generally proportional to $L_{\mathrm{D}}=\sqrt{\epsilon
kT/[e^2(n+p)]}$, where $n$ and $p$ are the free electron and hole
concentrations, respectively.

\begin{figure}
   \centering
   \includegraphics[width=7cm]{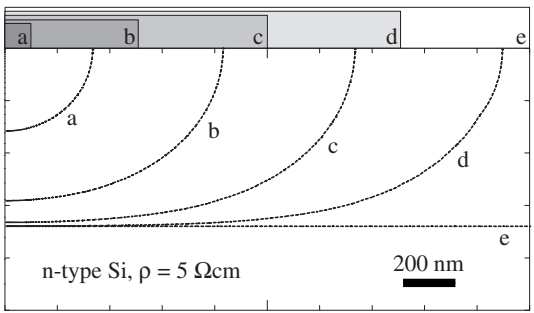}
   \caption{The dashed lines indicate the edge of the depletion region
   for various disc-shaped contacts (radii ranging from 100nm (a) to infinite (e)),
   taken from a numerical solution of the Poisson equation at 300~K.
   It clearly shows the size dependence of the depletion width for
   contact radii smaller than a few times $L_{\mathrm{D}}$ (which is 150~nm
   here). The left-hand vertical axis is an axis of rotational symmetry.
   The SBH is 0.67~eV.}
   \label{fig:depl}
\end{figure}

For diodes of finite size, the barrier can be much narrower. This
is because any charged object (here the metallic side of the
interface) with effective size $a$ at potential $V_0$ with respect
to infinity gives rise to a potential that drops as $V\approx
V_0\cdot (a/r)$ when the distance $r\gtrsim a$. Therefore the
barrier thickness will be at most a few times $a$, even in the
absence of free charge carriers. If there {\em are} free carriers
they can only make this barrier narrower---in the same way as with
the large diode---by depletion or accumulation close to the
interface. If, however, the interface size is much smaller than
$L_{\mathrm D}$, this additional screening can be completely
neglected. In the remainder of this paper, we use the terms
``large'' and ``small'' for diodes of which the interface size is
larger or smaller than $L_{\mathrm D}$, respectively. The
crossover to this new regime of small diodes is visualized in
Fig.~\ref{fig:depl}, with parameters similar to those in the
experiments. This figure is based on numerical solutions of the
Poisson-equation in silicon for various interface diameters.
The main consequence for electrical transport is, that the narrow
barrier in small diodes can make tunneling the dominant transport
mechanism (instead of thermionic emission) even at very low doping
levels.

Note that our description of small diodes has some similarity to
that of SBH-inhomogeneities in large diodes as analyzed by Tung
\cite{tung92}. There, the effect of small patches with lower SBH
on the space charge region is found to extend for only a few times
the size of these patches.

\begin{figure}
   \centering
   \includegraphics[width=8cm]{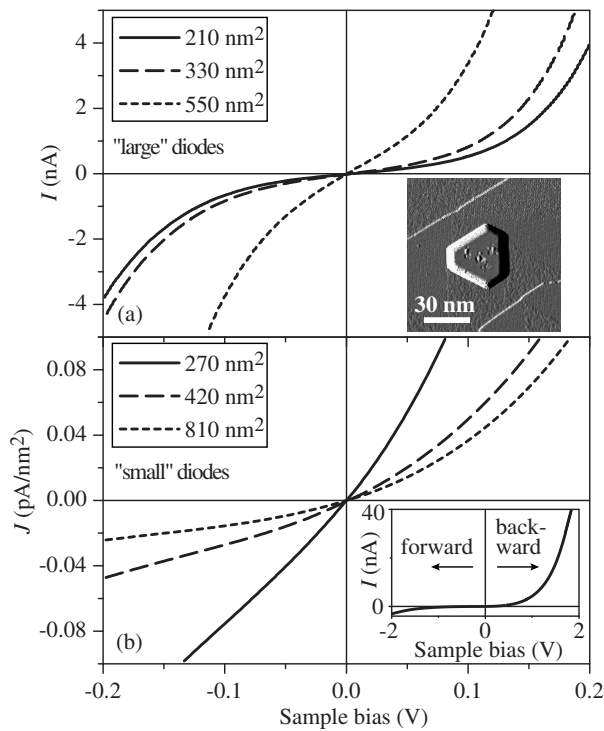}
   \caption{(a) Measured $IV$-curves for various island sizes on the
   $0.01\,\Omega$cm substrate. The zero-bias conductance
   is positively correlated to the island size. The inset shows a
   differential STM-image of a typical island after several
   $IV$-measurements.
   (b) Measured average current \emph{density} for various island sizes
   on the $10\,\Omega$cm substrate. Smaller islands have larger
   current densities due to the narrower barrier. The insert
   shows the full $IV$-curve for the 810~nm$^2$ island.}
   \label{fig:lowhigh}
\end{figure}

All experiments were performed in an UHV system with a base
pressure of $5\cdot 10^{-11}$~mbar. About 0.3 monolayers of Co
atoms were evaporated onto a clean, $7\times 7$-reconstructed
Si(111) surface which was held at room temperature. Then the
sample was subsequently annealed at $500\,^\circ\!$C and
$800\,^\circ\!$C, both for about 5~min, so that hexagon-shaped
epitaxial CoSi$_2$-islands were formed (Fig.~\ref{fig:lowhigh}(a),
inset) \footnote{It must be mentioned that the CoSi$_2$-islands
grow \emph{into} the substrate. However, the three occurring
CoSi$_2$/Si(100)-facets \cite{ilge00} do have the same SBH as the
CoSi$_2$/Si(111) interfaces \cite{tung84b,zhu00}.}. The height of
the islands ranged from 2--4~nm (with respect to the silicon
surface), while the diameters were in the range 15--30~nm. The
inter-island distances were much larger than the island diameters.
Next, the Si-surface reconstruction was destroyed by exposing the
surface to atomic hydrogen for 10~min, while the surface was held
at $400\,^\circ\!$C. The samples used were n-type doped with
resistivities of $10\,\Omega$cm and $0.01\,\Omega$cm,
respectively. After preparation, the surface was inspected with an
STM. The $IV$-measurements were done (at room temperature) by
positioning the STM-tip over an island and lowering it by a
distance $\Delta z$, sufficient to make contact to the island
(with feedback loop switched off). Then the current was measured
while ramping the voltage. The value of $\Delta z$ was determined
by lowering the tip at a fixed bias and measuring the current.
After the expected, initial exponential increase, the current
reached a constant, maximum value when it was lowered by $\sim
9\,$\AA. To ensure good contact, in all $IV$-measurements $\Delta
z = 15\,$\AA\ was used.

The $IV$-measurements did hardly deteriorate the imaging quality
of the STM-tip (Fig.~\ref{fig:lowhigh}(a), inset) and showed
excellent reproducibility when repeated on the same island. The
advantage of this type of measurement (as compared to the usual
STS) is that the measurements are not dominated by the properties
of the vacuum-gap, but instead it is possible to directly probe
the properties of the buried metal-semiconductor interface. The
electrochemically etched tungsten STM-tips were prepared to be
contaminant and oxide-free by in situ annealing and
self-sputtering with Ne. Since both CoSi$_2$ and the tip are
metallic, no barrier at the tip-island interface is expected.
Furthermore, the resistance of the back-contact on the sample was
measured to be at most a few k$\Omega$. Therefore, we can be sure
that we are really probing the properties of the island-substrate
interface.

From our measurements we learned that the nature of the surface
reconstruction (Si-$7\times 7$ or Co-induced $1\times 1$)
surrounding the islands greatly influenced the acquired
$IV$-curves, presumably because of surface conduction
\cite{hasegawa96} or environmental Fermi-level pinning resulting
in additional band bending. For this reason the surface
reconstruction was first destroyed by exposing it to atomic
hydrogen, as described before. This procedure always led to a
decreased conductance, apparently cutting off a surface-related
transport channel.

Although the size of our islands is limited to approximately
15--30~nm (due to the used growth mode), both the small and the
large diode regime can be addressed by varying the doping level.
Indeed, on the $10\,\Omega$cm substrate, where the doping level
predicts a screening length of $\sim 1\,\mu$m, we are far into the
small-diode regime. On the $0.01\,\Omega$cm substrate (screening
length $\sim 10$~nm) we are just at the other side of the
crossover.

It turned out to be impossible to fit our $IV$-data of the small
diodes to the standard diode equation $I\propto[\exp(eV/kT)-1]$,
even at small bias. This is a first hint that the dominant
transport process is not thermionic emission, here. A further
clear indication of the special behavior of small diodes is the
much higher conductance than expected from downscaling a
conventional diode. The inset of Fig.~\ref{fig:lowhigh}(b) shows a
typical $IV$-curve acquired on a small diode ($10\,\Omega$cm
substrate). The specific contact resistance $R_{\mathrm{c}}$ (the
zero-bias differential resistance multiplied by the island area)
for this measurement is $1\cdot 10^{-2}\,\Omega$cm$^2$, which is
$\sim 10^4$~times lower than for conventional diodes with a
barrier height of $0.67$~eV \cite{sze81,zhu00}. Considering the
facts that in a conventional diode on a $10\,\Omega$cm substrate
$J_{\mathrm{th}} / J_{\mathrm{T}}$ (ratio of thermionic and tunnel
current) is expected to be $\sim 10^{10}$ \cite{chang70} and that
$J_{\mathrm{th}}$ is independent of the barrier thickness, the
total current increase requires an increase of the tunnel current
by a striking factor of $\sim 10^{14}$. Such a large increase can
reasonably well be explained by a considerably reduced barrier
thickness. Besides, pure thermionic emission would lead to a
saturation current at positive sample bias of approximately
$1\cdot 10^{-7}$~nA for this SBH \cite{sze81}, while the observed
current was much larger. This also indicates the presence of an
important, additional conduction path. Note that it is not
necessary to assume a \emph{lower} SBH to explain our data.

To further test our hypothesis, the dependence of the (small-bias)
conductance on the island area was measured. The large diodes
($0.01\,\Omega$cm substrate) behaved as expected: the barrier
thickness is fixed by $L_{\mathrm{D}}$, so that a larger diode
area leads straightforwardly to a larger conductance (see
Fig.~\ref{fig:lowhigh}(a)). The observed scaling is not linear,
which is presumably due to the contribution of the edges, which
are not included in our simple model. The measurements in
Fig.~\ref{fig:lowhigh}(b) show that in small diodes the
conductance \emph{per unit area} decreases with increasing diode
area. This is fully consistent with our model, which predicts a
thicker barrier for larger interfaces.

Finally, we want to mention the behavior of small diodes at large
bias. In conventional diodes, the current saturates at reverse
bias. At forward bias, after the initial exponential increase, the
current is limited by the serial resistance of the bulk
semiconductor. In small diodes, the situation is completely
different, so that it might even reverse the expected diode
operation (see Fig.~\ref{fig:lowhigh}(b), inset, where the current
at forward (negative) bias is smaller than at reverse bias)
\footnote{It has been reported that a significant amount of
acceptor-like impurities can be incorporated in the top layer of
the substrate during sample flashing at $1200\,^\circ$C in UHV
\cite{liehr87}, effectively reversing the doping to p-type in
(initially) n-type low-doped samples. Nevertheless, the
concentration of these possible p-type dopants is expected to be
so low that it does not affect our main arguments.}. We found that
the observed behavior can be explained qualitatively by
considering the small diode as a ballistic point contact and
furthermore by taking into account Fowler-Nordheim tunneling from
the diode's edges at positive sample bias.

In conclusion, we have measured electrical transport through
epitaxial nanometer scale metal-semiconductor interfaces. Both the
observed high zero-bias conduction and the dependence of the
zero-bias conduction on the diode area support our model for the
extent of the space charge region for interface sizes smaller than
the free carrier screening length. This phenomenon provides a way
to tune the Schottky barrier thickness lithographically for a
fixed doping concentration, which can be useful in making tunnel
contacts.

We wish to thank J.~Caro for detailed discussions concerning this
work and W.J. Eijsenga and W. Crans for granting us access to
their ``Avant! Medici'' device simulation software. One of us,
S.~R., wishes to acknowledge fellowship support from the Royal
Netherlands Academy of Arts and Sciences.

\end{document}